# Topological Floquet bound states in the continuum

Chunyan Li,[1,2,*] Yaroslav V. Kartashov,[2] and Vladimir V. Konotop[3]

[1]*School of Physics, Xidian University, Xi'an, 710071, China*
[2]*Institute of Spectroscopy, Russian Academy of Sciences, 108840, Troitsk, Moscow, Russia*
[3]*Departamento de Física and Centro de Física Teórica e Computacional, Faculdade de Ciências, Universidade de Lisboa, Campo Grande, Ed. C8, Lisboa 1749-016, Portugal*
*Corresponding author: chunyanli@xidian.edu.cn*

**A honeycomb array of helical waveguides with zigzag-zigzag edges and a refractive index gradient orthogonal to the edges may support Floquet bound states in continuum (BICs). The gradient of the refractive index leads to strong asymmetry of the Floquet-Bloch spectrum. The mechanism of creation of such Floquet BICs is understood as emergence of crossings and avoided crossings of the branches supported by spatially limited stripe array. The whole spectrum of a finite array is split into the bulk branches being continuation of the edge states in the extended zone revealing multiple self-crossings and bulk modes disconnected from the gap states by avoided crossing. Nearly all states in the system are localized due to the gradient, but topological edge states manifest much stronger localization than other states. Such strongly localized Floquet BICs coexist with localized Wannier-Stark-like bulk modes. Robustness of the edge Floquet states is confirmed by their passage through a localized edge defect in the form of a missing waveguide.** © 2022 Optica Publishing Group

In the celebrated work [1] of J. von Neumann and E. Wigner it is proven that certain states in the continuum can be localized and decoupled from the extended modes. Different mechanisms of formation of such bound states in the continuum (BICs) were suggested, including mechanisms prohibiting coupling to continuum states due to different symmetry, interfering resonances, separability of underlying Hamiltonian in multidimensional settings, accidental BIC formation and many others, as described in reviews [2-4] and theoretical proposals [5-9]. In photonics, BICs have been suggested and observed experimentally in diverse systems including photonic crystals and slabs [10-13], arrays of waveguides or dielectric cylinders coupled to defect channels [14-16] or featuring engineered coupling constant profiles [17,18], anisotropic optical crystals [19], polaritonic microcavities [20] and lasers [21].

A new type of BIC was recently suggested in periodically driven Floquet system. Such Floquet BICs appear in the bulk of ac-driven by sinusoidal field one-dimensional lattice around frequencies, at which dynamic localization occurs due to collapse of the quasienergy band [22] (such BICs may coexist with standard bound states outside continuum arising due to periodic driving [23]). One-dimensional Floquet BICs residing at the edges of finite lattices with time-modulated coupling constants and linear gradient have been reported in [24], while two-dimensional Floquet BICs were obtained in [25]. All these states were encountered in topologically trivial systems. In this Letter we introduce a new topological mechanism of formation of Floquet BICs and show that such states may form in two-dimensional waveguiding system, where longitudinal modulations are responsible for nontrivial topology.

Topological insulators are materials behaving as conventional insulators in their bulk, but at the same time admitting topologically protected localized states at their edges. Unidirectional in-gap edge states in systems with broken time-reversal symmetry can propagate through surface defects without backscattering and do not couple to bulk modes. Recent progress in realization of photonic topological insulators is described in [26,27]. Floquet insulators occupy among them a special niche, since nontrivial topological phases arise in them due to modulations of the structure in evolution variable (time or propagation direction) [28,29], and enable observation of intriguing phenomena at optical frequencies in linear regime, including anomalous topological edge currents [30,31].

In this Letter, we obtain a new type of the spectrum of a Floquet topological insulator constructed of a helical waveguide array with a gradient of the linear refractive index directed transversely to a zigzag-zigzag edge of the array [see schematics in Fig. 1(a)]. Such a system features two edge states in the gap between the Dirac points. Even a small refractive-index gradient introduces asymmetry of the spectrum. Increase of this gradient results in penetration of the branch of the edge modes into the spectrum of the bulk modes [Fig. 1(b)-(i)] thus creating Floquet BICs. In an array bounded in the $x$-direction, the appearance of such BICs can be understood as crossing and avoided crossing of spectral branches. The found topological unidirectional pseudo-BICs are immune to defects.

Propagation of light in the above array [Fig. 1(a)] is governed by the Schrödinger equation:

$$i\partial_z\psi = \mathcal{H}\psi - Gx\psi, \quad \mathcal{H} = -(1/2)\nabla^2 - \mathcal{R}(x,y,z) \qquad (1)$$

where $\psi$ is the dimensionless amplitude of the light field; $\nabla=(\partial_x,\partial_y)$, $x,y$ and $z$ are scaled [23] transverse and longitudinal coordinates; the function $\mathcal{R}(x,y,z)=p\sum_{mn}e^{-[\mathbf{r}-\mathbf{r}_{mn}-\mathbf{s}(z)]^2/d^2}$ describes honeycomb array composed from helical waveguides of width $d$, with spacing $a$, depth $p$ and centered at $z=0$ in the points $\mathbf{r}_{mn}$; the function $\mathbf{s}(z)=r_0[1-\cos(\omega z),-\sin(\omega z)]$ sets waveguide rotation law with radius $r_0$ and period $Z=2\pi/\omega$; $G$ is a small refractive index gradient that can be emulated by additional waveguide bending in the transverse plane. The array is $y$-periodic with period $Y=3^{1/2}a$ and $z$-periodic with period $Z$. Mathematically the Eq. (1) is equivalent to the Schrödinger equation describing a quantum particle in a potential, with time emulated by the propagation coordinate $z$. Thus, the eigenmodes of the Eq. (1). $\psi_{\nu k}(x,y,z)=u_{\nu k}(x,y,z)e^{ibz+iky}$, are the Floquet-Bloch waves with quasi-propagation constants $b\in[-\omega/2,+\omega/2]$ defined $\mathrm{mod}\,\omega$, Bloch momenta $k$ along the $y$-axis in the first BZ $k\in[-\mathrm{K}/2,+\mathrm{K}/2]$, $\mathrm{K}=2\pi/Y$, and $u_{\nu k}(x,y,z)=u_{\nu k}(x,y+Y,z)=u_{\nu k}(x,y,z+Z)$, where $\nu$ denotes band index. In the $x$-direction, the array is finite, but sufficiently large, namely containing at least nine cells [each with six waveguides, see Fig. 2(b)].

The waveguide helicity breaks time-reversal symmetry $\mathcal{T}$ with respect to some instant $z_0$; it is defined by $\mathcal{T}\psi(x,y,z-z_0)=\psi^*(x,y,z_0-z)$, where asterisk stands for complex conjugation. Indeed, $\mathcal{R}(x,y,z-z_0)\neq\mathcal{R}(x,y,z_0-z)$ for any $z_0$. Furthermore, in the rotating frame $(\tilde{x},\tilde{y})=(x,y)-\mathbf{s}(z)$, using the ansatz $\phi=\psi\exp[-iG\tilde{x}z+(i/2)\int_0^z\mathbf{A}^2(z')dz']$, Eq. (1) is transformed into the Schrödinger equation: $i\partial\phi/\partial z=-(1/2)[\tilde{\nabla}+i\mathbf{A}(z)]^2\phi-R(\tilde{x},\tilde{y})\phi$ with the synthetic magnetic field $\mathbf{A}=Gz\hat{i}-\partial\mathbf{s}/\partial z$ and $\tilde{\nabla}=(\partial_{\tilde{x}},\partial_{\tilde{y}})$. This field lifts the degeneracy of two edge states at the opposite zigzag edges in the Bloch momentum interval $\mathrm{K}/3\leq k\leq 2\mathrm{K}/3$, opens a topological gap in the spectrum, and makes the above edge states unidirectional, as shown in Fig. 1(b) for $G=0$. Meantime, at $G=0$ the $\mathcal{P}_x\mathcal{T}$-symmetry, where $\mathcal{P}_x\psi(x-x_0,y,z)=\psi(x_0-x,y,z)$, remains, i.e. $[\mathcal{H},\mathcal{P}_x\mathcal{T}]=0$ for properly chosen $x_0$ (depending on the choice of the system of coordinates). The $\mathcal{P}_x\mathcal{T}$-symmetry imposes the parity symmetry of the Floquet-Bloch spectrum in Fig. 1(b) where the branches of the edge states cross at $k=\pm\mathrm{K}/2$ in the middle of the gap. Indeed, if $\psi_{\nu k}(x-x_0,y,z-z_0)$ is an eigenfunction at $G=0$, then $u^*_{\nu,-k}(x_0-x,y,z_0-z)e^{ib(z-z_0)-iky}$ is the eigenfunction too.

Positive gradient in the $x$-direction breaks the $\mathcal{P}_x\mathcal{T}$-symmetry of the system and symmetry of the Floquet-Bloch spectrum $b(k)$ leading to its transformation with increase of $G$ illustrated in Fig. 1(c)-1(i). Now the crossing point of edge states is shifted towards the former Dirac point at $2\mathrm{K}/3$ [Fig. 1(c)-1(f)]. Obviously, the spectrum remains $\mathrm{K}$-periodic. Reversing the sign of gradient $G$ shifts the intersection point in the $k$-space in the opposite direction, towards former Dirac point at $\mathrm{K}/3$ [compare Fig. 1(g) and (f)]. For a certain $x$-size of the array the total gap disappears for sufficiently large gradients $G$ [Fig. 1(h)]. Increasing the number of cells in the $x$-direction at fixed $G$ leads to broadening of the bands (it occurs because the gradient effectively reduces $b$ values for modes with maxima closer to the left edge and increases $b$ for modes with maxima near the right edge), their eventual overlap and disappearance of the total gap in the longitudinal BZ [see Fig. 1(i) calculated for 18 cells], but even in this case branches of states localized at the edges are clearly resolvable. Comparison of Figs. 1(i) and 1(f) shows that crossings of these branches with bulk states occur at different values of $k$ (and $b$) depending on the array size.

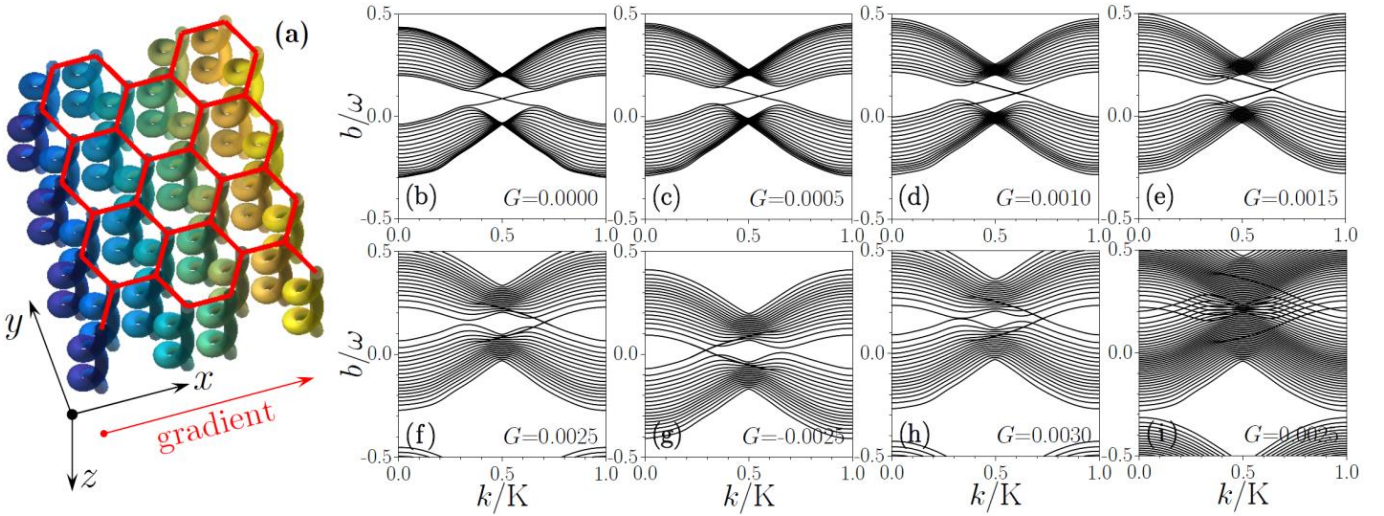

Fig. 1. (a) Schematic 3D representation of a honeycomb array of helical waveguides. Gradually varying color represents linear gradient of the refractive index. (b)-(i) Quasi-propagation constant *vs* normalized Bloch momentum for different values of the gradient $G$ indicated in the panels. In (b)-(h) the array contains 9 cells in the $x$-direction, in (i) – 18 cells. Here and in all figures below $r_0=0.7$, $a=1.7$, $d=0.4$, $p=12$, $Z=8$.

Simultaneously with growing asymmetry of the spectrum due to the gradient, we observe shifts of the topological edge states into the bulk bands. To understand this, in Fig. 2(a)-(c) we employ the fact that the structure is finite along the $x$-direction and therefore its spectrum contains a finite number of eigenvalues defined by the number of waveguides in the unit cell (i.e., it is quasi-continuous with a number of eigenvalues increasing with $x$-width of the structure).

At $G=0$ [Fig. 1 (b)], each of two edge branches continues at the bottom of the upper band and top of the lower band of the respective bulk spectrum. Employing the extended band scheme, $k\in[0,2\mathrm{K}]=\mathcal{I}_2$, one verifies that the edge branches are not independent, but represent continuation of each other: by following continuously one of them, after crossing two BZs, one returns to the starting point. When a weak gradient is applied, the spectrum deforms asymmetrically, but the connection of the edge states in the extended zone scheme $\mathcal{I}_2$ does not disappear immediately. Instead, one observes avoided crossing [see points 1 and 3 in Fig. 2(c)], where the edge mode approaches (couples with) a bulk mode.

Further increase of the gradient results in the transformation of the two avoided crossings into crossing points. Now, it is convenient to consider an extended zone scheme $k \in [0, 4\,\mathrm{K}] = \mathcal{I}_4$. Starting with a given edge state in the gap and following the branch over $\mathcal{I}_4$ one again returns to the starting point. Now the *two* bottom branches of the upper band and *two* top branches of the lower band represent the same spectral branch in the extended zone scheme. In Fig. 2(a) and 2(b) we present a close look at the situations, where upon further increasing of the gradient the avoided crossings of the gap edge branches with the bulk modes switches from the third [Fig. 2(a)] to the fourth [Fig. 2(b)] lowest bulk modes (shown by red arrow). In these cases, all branches between the avoided crossing points (only the upper ones are shown) represent a unique continuous branch in the extended $\mathcal{I}_6$ and $\mathcal{I}_8$ zone schemes, respectively. Thus, one can distinguish two types of the bulk modes: those continuously connected with the edge states in the respective extended $\mathcal{I}_n$ zone and manifesting self-crossing in the reduced BZ, and bulk modes disconnected by avoided crossing from the edge states. Upon increase of $G$ new bulk modes become continuously connected to the edge states until avoided crossing again occurs between the gap branches and the edges of the bulk spectrum [points 1 and 3 in Fig. 2(c)]. Meantime, upon increasing the size of the system along the $x$-direction the spectrum becomes more and more dense, leading in the continuous limit to the Floquet BIC.

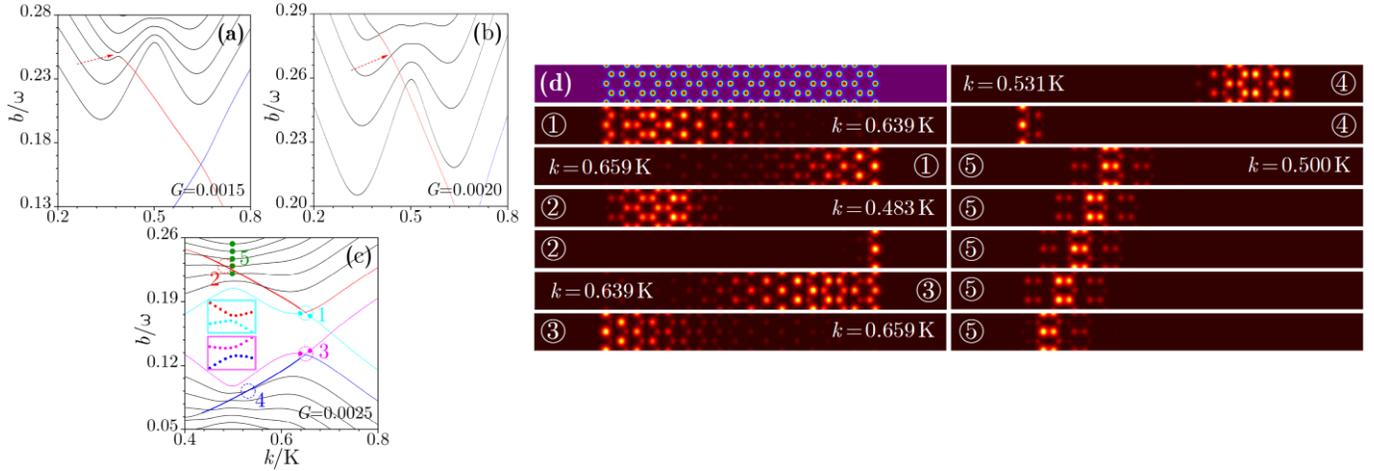

Fig. 2. (a)-(c) Zoom of Floquet-Bloch spectrum for different $G$ values indicated in the panels. The red arrows in (a) and (b) indicate an avoided crossing (a), which is transformed into crossing (b) when $G$ increases. In (c) two avoided crossings (groups 1 and 3, and the respective insets) and some of crossings (groups 2 and 4) are highlighted with open dotted and dashed circles, respectively. Red and blue colors indicate bulk branches separated by the avoided crossing from the topological edge branches shown by cyan and pink. Bold and thin color lines in (c) indicate strongly and weakly localized modes, whose structures for the chosen points are detailed in (d) where the field modulus distributions are shown. The respective Bloch momenta are indicated in the panels. Modes corresponding to solid green circles (group 5 at $k=0.500\,\mathrm{K}$) in (c) represent Wannier-Stark states having nearly equidistant quasi-propagation constants.

An essential difference of our system as compared to commonly known cases where BICs are observed, is that now nearly all modes are spatially localized due to nonzero gradient. Examples of the modes in the vicinity of the "critical" points are shown in Fig. 2(d), where we have chosen the case of $G = 0.0025$ [corresponding to the spectrum in Fig. 2(c)]. Analyzing the groups of modes 1 and 3, one observes mode transformations when passing the avoided crossing points. The modes [denoted by cyan and pink solid circles in Fig. 2(c)] at different sides of such points are localized at different edges of the array. This means that at certain value of the Bloch momentum (for example, between $k = 0.639\,\mathrm{K}$ and $k = 0.659\,\mathrm{K}$) the modes become fully extended over the whole stripe (not shown). Deviation of Bloch momentum from the avoided crossing value results in stronger localization of the modes. Particularly strong localization is observed for the modes, which appear as pseudo-continuation of the edge branches inside the gap [thick red and blue lines disconnected by avoided crossing from the respective cyan and pink lines in Fig. 2(c)]. Indeed, in the crossing points (groups 2 at $k = 0.483\,\mathrm{K}$ and 4 at $k = 0.531\,\mathrm{K}$) one observes different localization of the modes belonging to the bulk branch in the reduced BZ and modes belonging to the pseudo-continuation of the topological gap modes (those crossing points correspond to the same branch in the extended zone scheme). No coupling occurs between edge and bulk modes in such crossing points.

Bulk modes are localized not only at the edges of the stripe, but can be centered at some point inside the stripe. A remarkable example of this situations is given by the set of the modes at the boundary of the BZ, $k = \mathrm{K}/2$. In this point quasi-propagation constants in the bulk bands become nearly equidistant, forming the Wannier-Stark ladder in the Floquet system. Some of these modes corresponding to the green circles in Fig. 2(c) are presented in Fig. 2(d). These modes are well-localized in the bulk; they are practical replicas of each other (except for modes adjacent to the left and right boundaries and coexisting for a given momentum $k$ with edge states) shifted by $3a/2$ horizontally and by $3^{1/2}a/2$ vertically. At $k \neq \mathrm{K}/2$ the bulk modes are not equidistant and substantially more extended.

To further prove that the obtained edge states are indeed Floquet BICs decoupled from the bulk modes, while still exhibiting topological protection, we studied interaction with defects in the form of missing waveguides of the edge states exactly corresponding to the intersection points with bulk modes (i.e., for states residing in the bulk bands rather than in the gap). In all cases a wide Gaussian $y$-envelope of width $w=25$ was superimposed on the input edge state. Propagation dynamics for edge states corresponding to the intersections highlighted by the circles 2 and 4 in Fig. 2(a) are shown in Figs. 3(a) and 3(b), respectively. In Fig. 3(a) the edge state collides with a defect when moving in the positive direction of the $y$-axis, while in Fig. 3(b) the edge state encounters defect propagating in the negative direction of the $y$-axis. In both cases no backscattering or considerable radiation into the bulk is seen despite the presence of bulk states with the same quasi-propagation constant values in the spectrum. Similar behavior was encountered for other intersections.

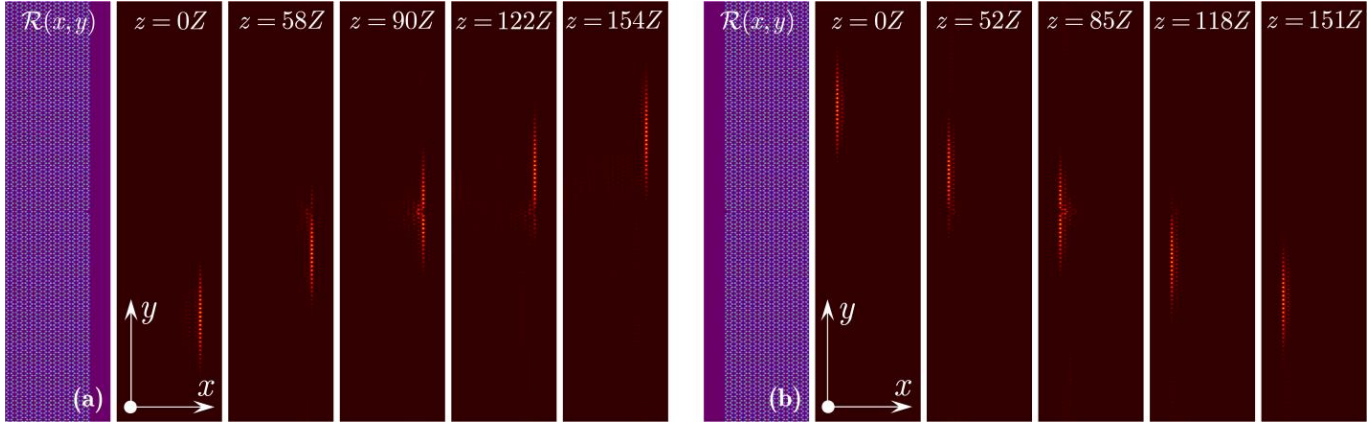

Fig. 3. Passage of the topological edge states with $k=0.483\,\text{K}$ (a) and $k=0.531\,\text{K}$ (b), indicated respectively by the circles 2 and 4 in Fig. 2, through the edge defect in the form of a missing waveguide. At $z=0Z$ the Gaussian $y$-envelope of width $w=25$ was superimposed on the input edge states.

In conclusion, we have reported on unusual transformation of the Floquet-Bloch spectrum of helical waveguide array due to the presence of linear refractive index transverse to the edge of the structure. Such gradient leads to notable asymmetry of the edge spectral branches and their shift into bulk bands, that nevertheless do not lead to coupling with bulk modes, at least for small gradients considered here.

**Funding:** National Natural Science Foundation of China (NSFC) (11805145); China Scholarship Council (CSC) (202006965016); Portuguese Foundation for Science and Technology (FCT) under Contracts PTDC/FIS-OUT/3882/2020 and UIDB/00618/2020.

**Disclosures:** The authors declare no conflicts of interest.